# Hybrid Exciton-Plasmon-Polaritons in van der Waals Semiconductor Gratings

Huiqin Zhang,[1] Bhaskar Abhiraman,[1,2] Qing Zhang,[3] Jinshui Miao,[1] Kiyoung Jo,[1] Stefano Roccasecca,[2] Mark W. Knight,[4] Artur R. Davoyan,[5] Deep Jariwala[1]*

[1]Department of Electrical and Systems Engineering, University of Pennsylvania, Philadelphia, PA 19104, USA
[2]Department of Physics, University of Pennsylvania, Philadelphia, PA 19104, USA
[3]Department of Electrical and Computer Engineering, National University of Singapore, Singapore 117583, Singapore
[4]Northrop Grumman Corporation 1 Space Park Drive, Redondo Beach, CA 90278, USA
[5]Department of Mechanical and Aerospace Engineering, University of California, Los Angeles, CA 90095, USA

*Corresponding Author: dmj@seas.upenn.edu

**Abstract:** Van der Waals materials and heterostructures manifesting strongly bound room temperature exciton states exhibit emergent physical phenomena and are of a great promise for optoelectronic applications. Here, we demonstrate that nanostructured multilayer transition metal dichalcogenides by themselves provide an ideal platform for excitation and control of excitonic modes, paving the way to exciton-photonics. Hence, we show that by patterning the TMDCs into nanoresonators, strong dispersion and avoided crossing of excitons and hybrid polaritons with interaction potentials exceeding 410 meV may be controlled with great precision. We further observe that inherently strong TMDC exciton absorption resonances may be completely suppressed due to excitation of hybrid photon states and their interference. Our work paves the way to a next generation of integrated exciton optoelectronic nano-devices and applications in light generation, computing, and sensing.

## INTRODUCTION

Control of light-wave dispersion at the nanoscale is of great importance for applications ranging from lasing to sensing to computing. Plasmonic and high contrast dielectric nanostructures(*1-4*) are a frontier approach for dispersion engineering (Fig. 1. A,B) at the dimensions comparable and smaller than the light wavelength. Materials with resonant quantum confined states provide an alternative route for controlling light propagation and interaction. This strategy has been effectively employed in a wide variety of materials ranging from inorganic III-V epitaxial quantum wells to organic small molecules and polymers, to even carbon nanotubes, and most recently, two-dimensional (2D) transition metal dichalcogenides (TMDCs) (Fig. 1. C). Due to weak intermolecular bonding or physical confinement, excitons in this class of materials exhibit strong binding energies and dominate optical responses even at room temperatures.(*5*) Among these excitonic materials, the 2D TMDCs of Mo and W are of particular interest since the strong exciton binding manifests in high refractive indices and extinction coefficients (k) near the exciton resonance. Light-exciton interaction has been extensively studied in monolayer direct-band gap TMDC films(*6-10*). These studies have involved coupling the monolayers to extrinsic plasmonic or dielectric optical meta-elements or cavities or metasurfaces (*11*). Strong coupling regimes and families of hybrid exciton modes have been observed. Nevertheless, light-interaction with monolayer films is challenged by the very large disparity between optical wavelength and film thickness. Recent demonstration of enhanced resonant light interaction with few layer thick TMDCs suggests that thin systems may be well suited for light dispersion (*5, 12, 13*) Here, we show that nanostructured multilayer TMDCs provide a natural platform for taming light-exciton interaction. Specifically, within a single integrated platform we observe multipartite interaction and hybridization, leading to strong coupling between three resonant excitations with an avoided crossing of 410 meV at room temperature. We further reveal a novel regime of light-TMDC interaction associated with suppression of inherent exciton absorption resonances due to interference of hybridized states at the nanoscale.

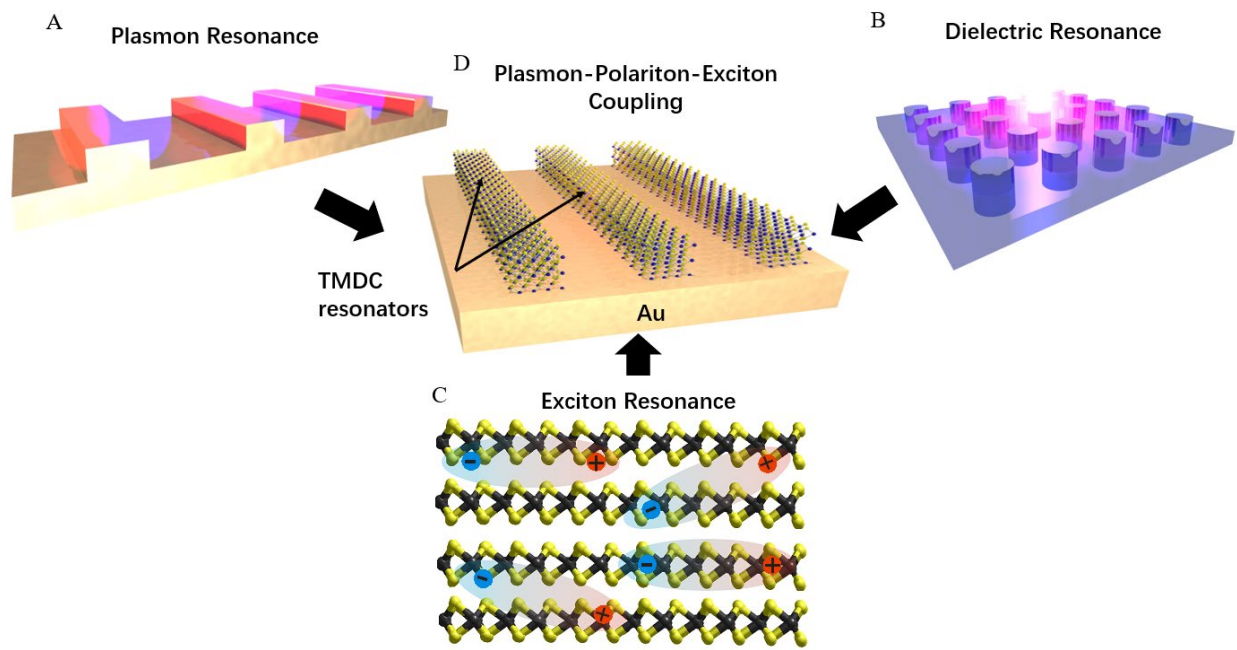

**Fig. 1. Conceptual illustration of multipartite light-materials interaction within nanoscale materials and patterned structures.** Light interaction with nanostructured metals (**A**) and dielectrics (**B**) gives rise to the excitation of plasmonic and polaritonic resonances, respectively. Dispersion of these resonances is determined mainly by the geometry of the system. (**C**) Reduced dimensional materials provide a different approach to strong light-materials interaction. Here, sharp optical resonances manifest in excitation of long-lived bound electron-hole pairs, with layered two-dimensional transition metal dichalcogenides (TMDCs) as prominent examples. (**D**) By mixing engineered geometric dispersion in metal-dielectric nanostructures with intrinsic excitonic resonances in TMDCs, a multifaceted, strongly-interacting interplay of photonic and electronic states can be achieved.

**RESULTS AND DISCUSSION**

We begin by systematically exploring optical reflectance spectroscopy, first in unpatterned (Fig. 2 A) and then in patterned cases (Fig. 2 B) of varying TMDC thickness. Our choice of the TMDC is $WS_2$, but our results are generalizable to other room temperature excitonic materials (see Supplementary Information S3). Micromechanically exfoliated $WS_2$ flakes were obtained on a template-stripped Au substrate. Upon exfoliation these flakes exhibit thickness-dependent colors resulting from Fabry-Perot-like resonances(*13*), resulting in translucent red, purple, blue, and white regions of crystal (Fig. 2 C). Periodic one-dimensional (1D) gratings were etched in single $WS_2$ flakes to increase the light-materials interaction strength and engineer optical dispersion. Structural characterization of grating etch patterns by atomic force microscopy (AFM) and scanning electron microscopy (SEM) (Fig. 2 C,D) shows deep-subwavelength $WS_2$ flake thicknesses, along with gratings etched through the full flake. The etching process results in some edge roughness, but these non-perturbative structural deviations do not modify the optically critical parameter, which is the grating period. Detailed optical micrographs, AFM profiles and SEM images are shown in supplementary S1.

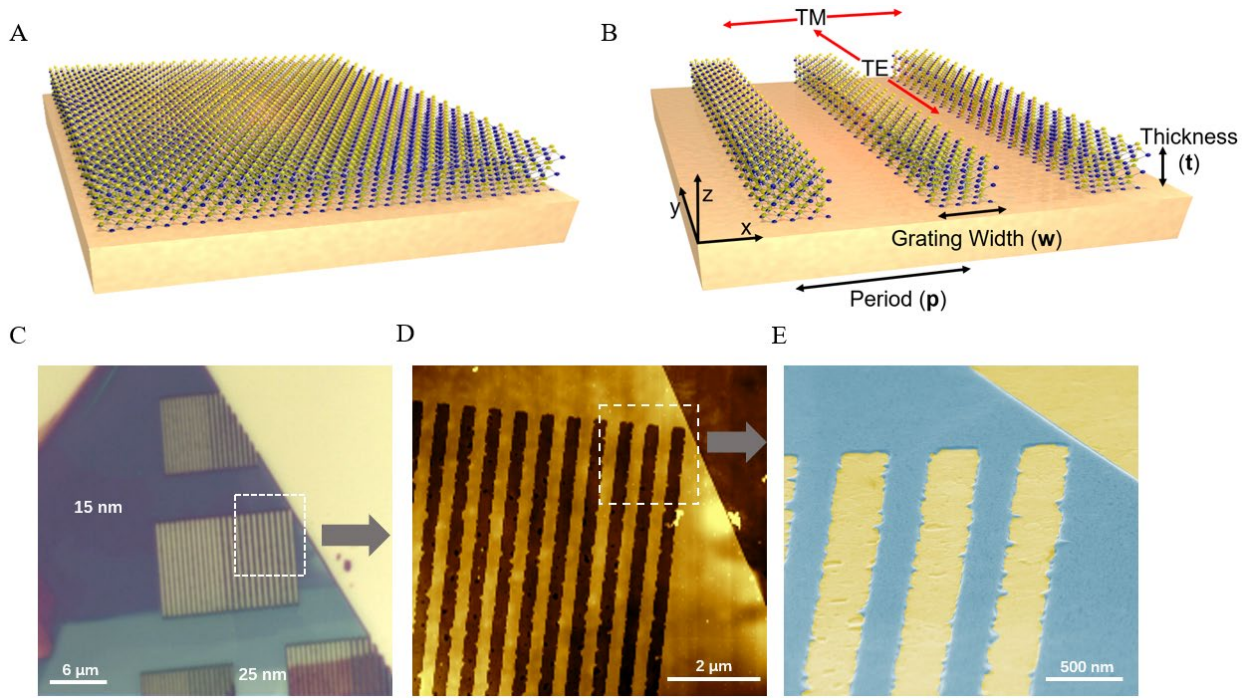

**Fig. 2. Nanopatterned multi-layer $WS_2$ grating resonators on gold.** (**A**) Schematic of an unpatterned multi-layer $WS_2$ flake on an Au substrate. (**B**) Schematic of a multi-layer $WS_2$ grating structure on a gold substrate. Red arrows denote the incident light polarization: TM polarization is defined such that the electric field is perpendicular to the grating, while the TE polarization electric field is parallel to the grating. Grate width (w), period (p) and thickness (t) are defined. (**C**) Unpolarized white light micrograph of $WS_2$ flake on Au substrate, showing patterned and unpatterned regions. Different colors appear due to different absorption spectra for different material thicknesses; 15 nm thick $WS_2$ exhibits a purple hue while 25 nm thick $WS_2$ appears pale blue. (**D**) Atomic Force Microscopy (AFM) image of the grating topography corresponding to dashed box of panel c. (**E**) False color Scanning Electron Microscopy (SEM) image of the boxed region in panel d acquired at 70° tilted angle. The grating structure with dimensions w = 300 nm and p = 600 nm is clearly seen.

Reflectance spectra for unpatterned $WS_2$ of varying thickness on an Au back reflector (Fig. 3 A,B) exhibit two reflectance dips (absorption resonance peaks). The primary exciton mode of $WS_2$ at

2.0 eV demonstrates high absorption independent of TMDC thickness. The simulated magnetic field profile indicates strong exciton driven light absorption, suggesting Beer-Lambert like absorbance, typical for bulk samples. In contrast, the magnetic field profile representing the Fabry-Perot resonance indicates a low-Q cavity mode formed by the combination of metal and high-index dielectric in the heterostructure. (*13, 14*) The cavity mode is observed for deep-subwavelength thicknesses, owing to the lossy dielectric of $WS_2$ on a reflective substrate facilitating near-unity absorptive resonances. The cavity mode couples to the primary exciton mode at the critical thickness ~15 nm, resulting in overall absorption enhancement. This coupling results in hybridization of the light (polariton) and matter (exciton) modes into exciton-polariton modes with a characteristic avoided crossing in the reflectance spectra. As the cavity mode is tuned through the exciton energy, the upper exciton-polariton (UEP) branch and lower exciton-polariton branch (LEP) are split.(*15*) At the point of avoided crossing of the exciton and the cavity polariton, the splitting energy is evaluated to be $\hbar\Omega$ = 170 meV by fitting the simulation data to a coupled oscillator model (Supplementary S2). Experimental reflectance spectra with varying thicknesses of $WS_2$ show remarkably good qualitative and quantitative agreement with the simulations (Fig. 3 B). Both the primary exciton peak and the weak B exciton peak ($\lambda$ ~ 515 nm) locations match well. Note that this strong coupling between a weak cavity mode and an exciton mode in the deep subwavelength thickness regime can be extended to other TMDC materials and geometries. Notably, such behavior cannot be observed in III-V semiconductors (*e.g.* GaP or GaAs) that possess similar band gaps but lack room-temperature excitonic features in their dielectric functions (see supplementary S3).

Periodic patterning provides an additional degree of control for light-exciton interaction and dispersion engineering (Fig. 3 C,D). Under TM polarization, we observe that several new absorption resonances emerge for a periodically patterned TMDC. Primarily, a resonant mode associated with a dielectric grating emerges at the highest-energies (~500 nm) while several plasmonic modes excited at lower energies (Fig. 3 E).(*16, 17*) Characteristic mode profiles are shown in the Fig. 3 E; different type of plasmonic mode localization and confinement is observed. For higher thicknesses (more than ~15 nm thick), these plasmonic modes further red-shift with increasing thickness. Detailed discussion on E and H field profiles of the modes is provided below and in supplementary information S4 and S5. Our simulation results closely match experimental reflectance spectra for three discrete thicknesses of 10, 14 and 20 nm (Fig. 3 D). Under TE polarization (Fig. 3 F), the simulated reflectance spectra behave similarly to the unpatterned case, implying that grating and plasmon modes are not excited, since there is no breaking of symmetry along TE polarization upon etching. Experimentally measured reflectance spectra also match the simulations for the TE case, and the spectra as well as mode profiles are shown in detail in supplementary S4. All simulations and experimental measurements are for TM polarization for the remainder of this work.

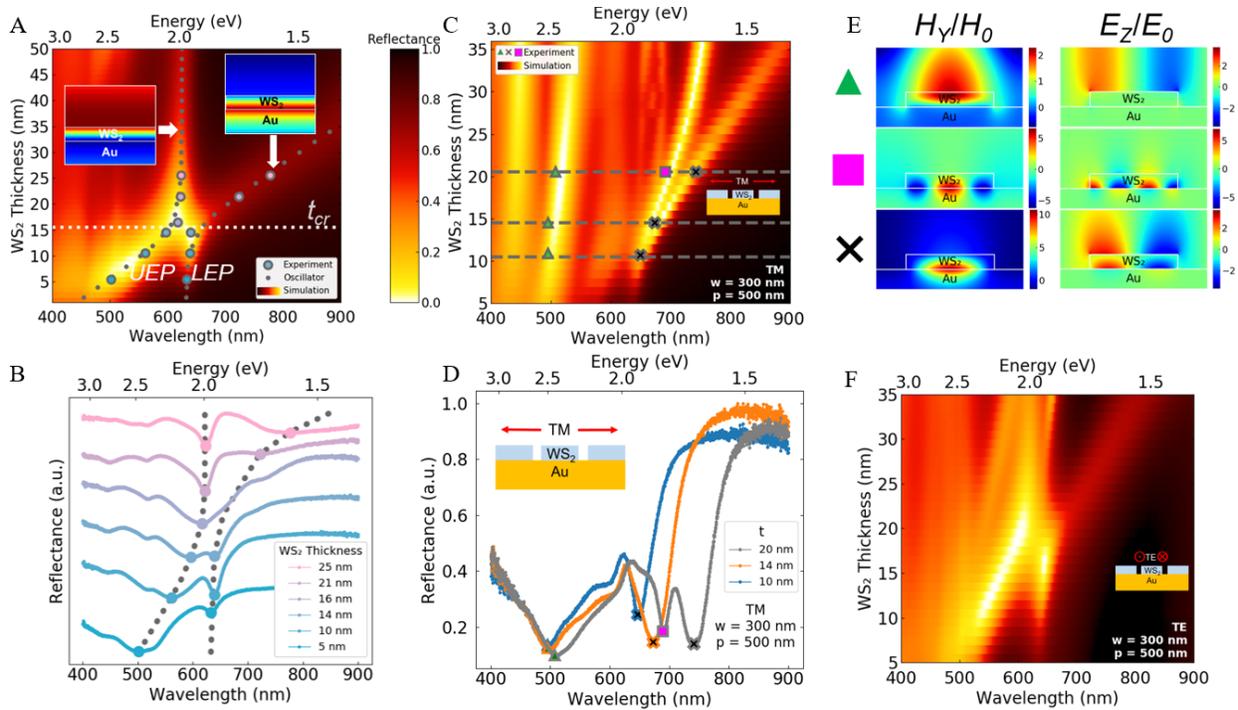

**Fig. 3. Below-gap resonant absorption in WS$_2$ on Au.** (**A**) Plot of simulated reflectance spectra of unpatterned WS$_2$ on Au with varying WS$_2$ thickness on the y axis. The primary exciton mode of WS$_2$ can be seen at ~2.0 eV which couples with the cavity mode strongly, splitting the exciton into the upper exciton-polariton (LEP) and the lower exciton-polariton (UEP). A low Q Fabry-Perot-like cavity is formed even in deep-subwavelength thickness regime as a result of the thin film interference. Insets show the H$_Y$ field profiles corresponding to the exciton mode (left) and the cavity mode (right). Reflectance color scale is the same for a, c and f. (**B**) Experimental unpolarized reflectance spectra of unpatterned WS$_2$ on Au with varying thickness. The experimentally measured thicknesses are 5, 10, 14, 16, 21, and 25 nm. The UEP and LEP peaks are emphasized with circles and are superimposed on panel a to show strong matching with simulation. (**C**) Simulated reflectance spectra under TM polarization of a 1D WS$_2$ grating on an Au substrate with varying thickness. The structure has a fixed width and period of 300 nm and 500 nm respectively. The non-dispersive mode induced by dielectric grating emerges as the highest-energy resonance (~500 nm) while plasmon resonances of varying orders emerge at lower energies. For higher thickness regimes, an additional plasmon resonance emerges which red-shifts with increasing thickness. (**D**) Experimental TM reflectance spectra of the 1D WS$_2$ grating on the Au substrate with w = 300 nm and p = 500 nm. The peaks corresponding to various modes are coded with matching symbols on c. (**E**) E$_Z$ and H$_Y$ mode profiles for the various modes matching with the corresponding symbols in c & d. (**F**) Simulated reflectance spectra under TE polarization of a 1D WS$_2$ grating on an Au substrate as thickness is swept. Experimental absorption peaks are marked for thicknesses of 10 nm, 14 nm, and 20 nm.

To understand further how the lateral patterning affects mode dispersion, we have investigated the dependence of reflection spectra on grating finger width. At the critical thickness (~15 nm), a plasmon-mode branch emerges from the LEP branch which suggests it has a hybrid character (Fig. 4 A). The experimental reflectance spectra for fixed t = 15 nm and p = 500 nm, and widths of 200 nm and 300 nm match well with the simulation. This along with the non-dispersive grating induced mode and a shallow exciton-polariton mode around 600 nm are all observable in the experimental spectra (Fig. 4 B). Beyond the critical thickness regime, more (higher order) plasmon modes emerge from the hybrid LEP mode (Fig. 4 C) which is once again verified in the experimental spectra, (Fig. 4 D) where a second peak emerges at 695 nm besides the first peak at 740 nm. Grating structures also offer a possibility of tuning frequency of these excited plasmons by changing grating geometrical parameters i.e. width. (Fig. 4 E). The grating with its own wave-vector presents momentum matching conditions to excite plasmons in integer multiples of if this wavevector. Higher-order plasmons (A,B,C in Fig. 4 E) emerge in the reflectance spectra as the width is increased, while the plasmon mode indicated by D ( Fig. 4 D) retains its place as the lowest energy branch. The $E_z$ and $H_y$ field profiles show plasmons of different orders trapped between the $WS_2$/Au interface (Fig. 4 F) verifying our assumption. Another interesting observation that we make in our system is the occurrence of hybrid plasmonic modes. At the crossing point of the plasmon mode labelled D in Fig. 4 E with one labelled B, there is a minor dip in intensity followed by re-emergence of the linear dispersion, which is slightly offset from the original SPR line. We designate these as hybrid modes retaining character of both the B and D plasmon branches. Extensive width dependence of the hybrid modes is shown in detail as an animation in Supplementary information S5. Note that, the plasmons observed in our system may be correlated to different types of plasmon modes (e.g., localized and surface plasmons). (1),owever, here due to the complex nature and multipartite interactions involved, we avoid making any such categorical classifications.

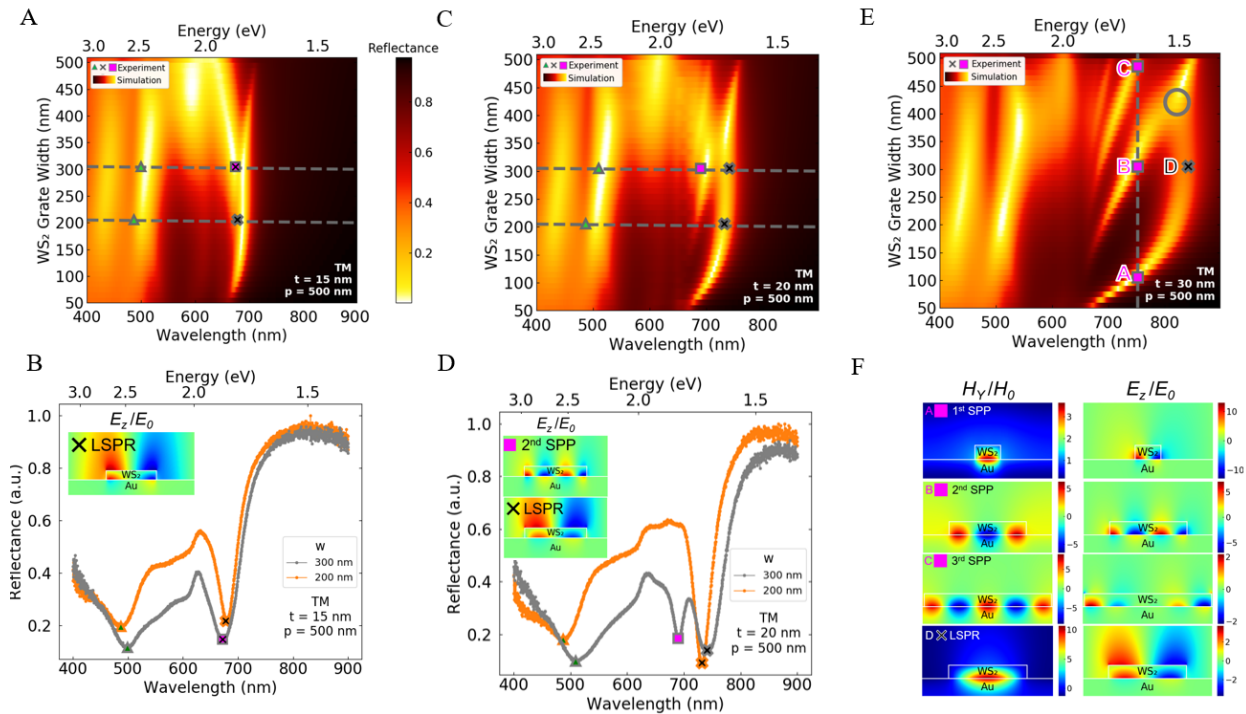

**Fig. 4. Plasmonic modes above critical-thickness of WS$_2$ on Au.** (A) Width dependence of simulated TM reflectance spectra. Period is fixed at 500 nm, and thickness is fixed at 15 nm

(critical thickness). The plasmonic peak emerges from the lower exciton-polariton (LEP) of the unpatterned case indicating a hybrid mode. Reflectance color scale is the same for a, c and e. (**B**) Experimental reflectance spectra of the WS$_2$ grating structure with t = 15 nm and p = 500 nm, and widths of 200 nm and 300 nm. Corresponding mode peaks are coded with matching symbols in a. Inset is the E$_z$ field profile of the first order plasmon mode. (**C**) Width dependence of simulated TM reflectance spectra. Period is fixed at 500 nm, and thickness is fixed at 20 nm (above critical thickness i.e. uncoupled regime). Since the thickness is sufficient to support cavity modes in the unpatterned case, plasmon modes (marked with square symbols) start emerging from the cavity mode. (**D**) Experimental reflectance spectra of the WS$_2$ grating structure with thickness and period fixed at 20 nm and 500 nm respectively. Spectra for widths of 200 nm and 300 nm are compared. For a width of 300 nm, a second-order plasmon mode emerges. Inset are the E$_z$ field profiles of the two plasmonic mode. (**E**) Width dependence of simulated TM reflectance spectra. Period is fixed at 500 nm, and thickness is fixed at 30 nm, way above the critical thickness regime. The WS$_2$ resonators provide index discontinuity for resonant SPR excitation. For a photon wavelength of 750 nm, the resonant widths are equal to 100 nm (A), 300 nm (B), and 470 nm (C), forming 1$^{st}$, 2$^{nd}$, and 3$^{rd}$ order SPR modes. The plasmon mode (D) stands at the lowest-energy resonance. The circle marked at the crossing point of the 2$^{nd}$ order plasmon mode (B) and (D) indicates a hybrid mode with characters of both. (**F**) E$_z$, H$_Y$ field profiles of the plasmonic modes labelled with corresponding points in e.

The above discussion summarizes the emergence of various plasmonic modes above critical thickness and their hybridization with each other. Thus far there has been no mention of the dielectric grating mode and its interaction with the exciton. This interaction is strong and notably observed for thicknesses below the critical value. The incident wave couples to the guided mode, and the dielectric-grating mode is formed under the momentum matching condition between the incident light-frequency (ω) and in-plane wave momentum (k$_x$) component generated by the 1D dielectric (WS$_2$) gratings.(*17, 18*) The dielectric grating induced mode couples the incident light in-plane which allows us to estimate a momentum vs frequency dispersion of the resonances. Below the critical thickness (t= 15 nm, or in the strong-coupling regime), the original primary exciton absorption peak is split into LEP (~630 nm) and UEP (~550 nm) modes as a result of strong coupling between the exciton and the cavity mode, as shown in Fig. 5 A for the case of 10 nm thick WS$_2$. Both the UEP and the LEP can further couple to the dielectric grating induced mode, forming a three oscillator coupling.(*19*) In a structure with a fixed width of 300 nm, the grating mode can be tuned by varying the period to be in resonance with both UEP and LEP modes. Upon that resonance, strong-coupling occurs among all three modes, leading to redshift of both UEP and LEP modes. The grating mode transforms into the upper polariton branch (UPB), while the lower polariton branch (LPB) switches from the LEP mode to plasmonic mode with increasing period. The UEP mode forms the middle polariton branch (MPB), which has a cut-off in the period range of 500 nm to 650 nm. A triple-oscillator model is applied to fit the system based on the assumption that the UEP branch is pumped at first as the driven force(Fig. 5 C).

$$\begin{pmatrix} E_d & g_{d-UEP} & g_{d-LEP} \\ g_{d-UEP} & E_{UEP} & g_{LEP-UEP} \\ g_{d-LEP} & g_{EP-UEP} & E_{LEP} \end{pmatrix} \begin{pmatrix} \alpha_1 \\ \alpha_2 \\ \alpha_3 \end{pmatrix} = E \begin{pmatrix} \alpha_1 \\ \alpha_2 \\ \alpha_3 \end{pmatrix}$$

(See supplementary 6 for in-depth discussion of the model and parameters). To calculate the splitting energy, we assume a near-zero value of $g_{d-LEP}$ so the splitting energy can be expressed in terms of the coupling strengths $g_{d-LEP}$ and $g_{UEP-LEP}$. From the experimental spectrum corresponding to the zero-detuning point (p = 600 nm), $g_{d-UEP}$ and $g_{UEP-LEP}$ are taken to be 116 meV and 89 meV respectively (Fig. 5 B). When the dielectric grating induced mode and the plasmonic mode are closest in energy, and the MPB is cut-off in between them, the giant avoided-crossing splitting energy is calculated to be $\hbar\Omega = 2(g_{d-LEP} + g_{LEP-UEP}) = 410\ meV$. Interestingly, the MPB cut-off is only replicated in the oscillator model when it is assumed that the MPB oscillator, i.e. the UEP mode, is the only one driven. This suggests the UEP mode is the primary excitation mechanism of this system. This cut-off of the MPB upon strong interaction between these three modes is fundamentally novel and unseen in earlier reports of strong light-matter coupling of quasiparticles in TMDCs.(20) Based on the electromagnetic simulation and experimental spectra (diamond symbol in Fig. 5 A & B), this cut-off of the middle branch, where the absorption reaches almost zero, suggests that the strong mode coupling results in an interference condition that prevents any electromagnetic field interaction with the lossy dielectric part. This is evident from the field line profile (Fig. 5 D). As compared with the UPB and LPB of absorption resonances where all the electric-field lines are recirculating or focused in the $WS_2$ resonators (Fig. 5 D triangle and cross), in the cut-off middle branch region (absorption dip), most electric-field lines are pushed out of or take the shortest part of interaction through the resonator, resulting in an electromagnetically 'invisible' resonator, qualitatively analogous to atomic systems(21, 22) (Fig. 5 D diamond). Such electromagnetic 'invisibility' is fundamentally new in any lossy photonic system let alone an excitonic, external cavity-less system. Lossy grating structures are not new and have been heavily investigated for Si, Ge and lossy metals based structures(23-27) for the purposes of light trapping and structural coloring. In all cases, electromagnetic wave simulations suggest field lines concentrating (focused) inside the lossy dielectric part(28) in contrast to our excitonic grating system where the field lines are pushed out of the lossy resonator part (Fig. 5 D).

The evolution of the strong coupling with thickness (Supplementary 7) and resonator width (Fig. 5 E) shows that there exists a critical range of resonator widths (50-330 nm) and a critical thickness (15 nm) under which this triple-oscillator coupling resulting in cut-off of the MPB is supported. The field profile of the high-Q plasmon mode (Fig. 5 E inset) shows significant leakage out of the TMDC, indicating sensitivity to the dielectric environment, with potential applications in chemical and biomolecular sensing(29) as well as coupling/tuning quantum emitters in other TMDCs or h-BN(30-32). Because the grating mode's field enhancement is localized at the top-corners of the resonator, the grating width is particularly critical for strong-coupling. For p = 600 nm and widths larger than 330 nm, since the location of the two modes are too separated, the grating mode lacks the momentum to strongly interact with the exciton, which can explain the fact that no more strong interaction occurs at the point where the grate width larger than 330nm. Additionally in the other hand, the strength of the polariton can be reduced when one dimension of the grating is limited. However, the LEP oscillator strength is much more robust compared to the pure excitons at widths lower than 300 nm (Fig. 5 E-F). In comparison to the strongly coupled case (t < 15 nm), the pure exciton absorption almost vanishes in the uncoupled regime (t > 15 nm) (Fig. 5 F). For these thicknesses of $WS_2$, the exciton does not couple to the cavity mode to form hybrid polaritons. The exciton mode at λ ~ 600 nm almost vanishes in strength before it can even interact with the grating mode due to its instability in the discontinuous $WS_2$ structure (Fig. 5 F). Mode B and C' correspond to the 2$^{nd}$ and 3$^{rd}$ order SPR modes while mode D corresponds to the plasmonic mode. This indicates that the exciton-polaritons (UEP and LEP modes) or hybrid particles have a much greater robustness and stability to strongly couple to a second photon as opposed to pure excitons.

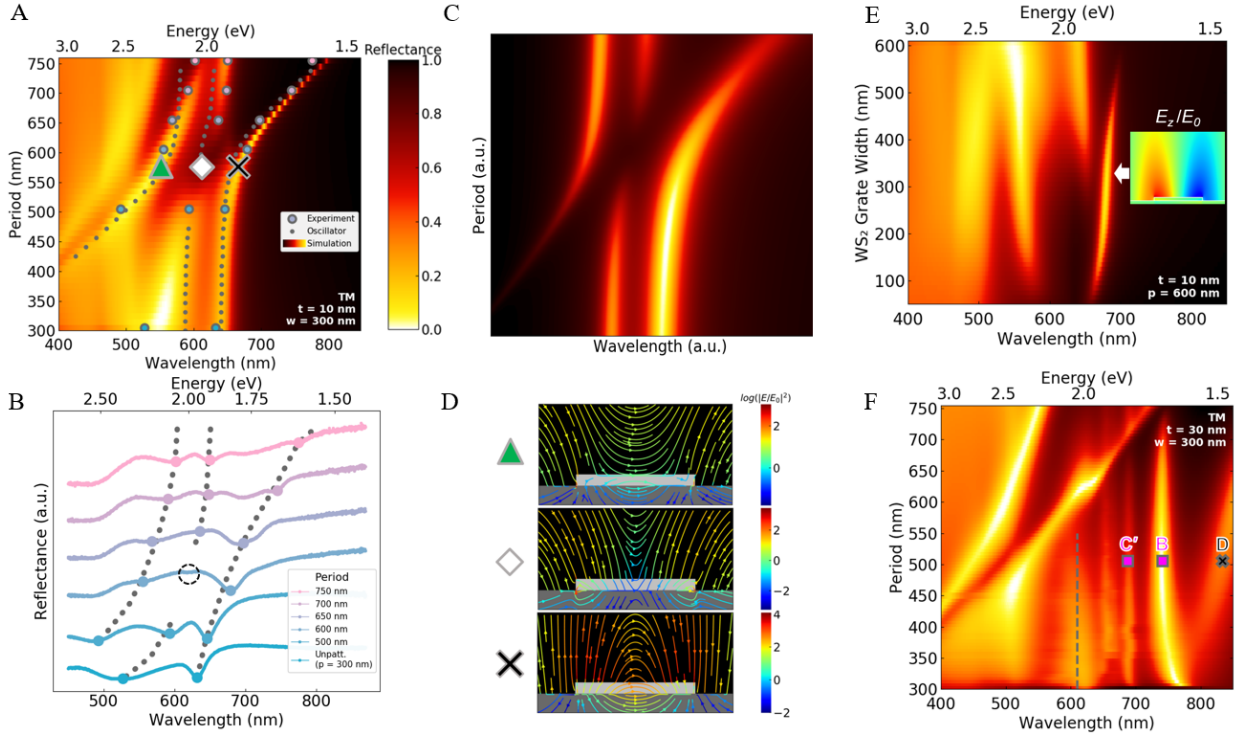

**Fig. 5. Strong three-oscillator coupling below critical thickness.** (**A**) Period dependence of simulated TM reflectance spectra. Thickness and width are fixed at 10 nm and 300 nm respectively. Period is swept from the unpatterned case (w = p = 300 nm) to a period of 750 nm. As period increases, the plasmonic peak shifts from 640 nm to 800 nm. The dashed lines represent the calculated absorption spectra based on the triple-oscillator model in Fig. 5 C. Reflectance color scale is the same for a, e and f. (**B**) Experimental reflectance spectra of the $WS_2$ grating structure with a fixed thickness of 10 nm and a width of 300 nm. Spectra for the unpatterned case, p = 500 nm, 600 nm, 650 nm, 700 nm, and 750 nm are plotted and offset for clarity. The marked peaks are also plotted in a, for comparison with simulation. (**C**) The calculated energy dispersion relation based on the triple-oscillator model. The three oscillators represent the grating mode, LEP mode and UEP mode. The cut-off of the middle polariton branch is accurately replicated with the coupled oscillator model. By fitting the oscillator model to the simulation and experiment data, we obtain the giant splitting energy of 410 meV. (**D**) Comparison of the field-line profiles of the UP, MP cut-off, and LP regions marked with triangles, diamonds and crosses respectively in Fig. 5 A and B. At the point marked with a diamond, almost all the electric field lines are repelled out of the resonator rendering the highly lossy $WS_2$ resonator "invisible" in the far-field. (**E**) Reflectance spectra of grating resonators showing the evolution of the strong-coupling-induced splitting as a function of width in the wavelength range of ~550-650 nm. Inset: the $E_Z$ field profile of the plasmonic mode for t = 10 nm. (**F**) Period dependence of reflectance of the WS$_2$ grating structure for t = 30 nm and w = 300 nm. The dashed line indicates the exciton mode at λ = 610 nm almost vanishes. Higher order plasmonic modes emerge from the cavity mode (which originates from the unpatterned case). The dielectric grating induced mode is unable to couple with the pure exciton without the UEP-LEP induced triple-oscillator coupling.

**CONCLUSIONS**

Our results highlight a previously unexplored aspect of TMDC excitons and optics, and opens a new regime in exploring strong light-matter interactions where the TMDC serves as a host for both excitonic and cavity modes, without the need for an external cavity. Prior studies have focused on either monolayers or very thick (>100 nm) TMDCs.(*12*) Our results show that in this intermediate thickness regime, there is possibility of combining multiple classes of light-matter interactions (plasmons, excitons and cavity polaritons) resulting in strong-coupling phenomena involving two photons or three oscillators resulting in observation of optical "invisibility" conditions(*33*). Further, our results showing a large mode-splitting energy (~410 meV) at room temperatures can help serve as a platform for exploring a rich diversity of polaritonic phenomena. While multilayer TMDCs are not emissive, they can serve as an underlying couplers for quantum point defects or other emitting gain media on top, potentially paving a way to exploring multi quantum point emitter entanglement, polariton condensation in molecular or quantum dot assemblies(*34, 35*) as well as low-threshold lasers at room temperatures.(*32, 34, 36*)

**MATERIALS AND METHODS**

**Sample fabrication**

Measurements were performed on a template stripped Au substrate(*13*). To ensure a flat surface of the gold back reflector rather than the rough surface of the evaporated gold, an epoxy-based peeling procedure is applied to the 100 nm-thick Au film evaporated on a clean polished Si wafer (parent wafer) using an e-beam evaporator (Kurt J. Lesker PVD 75). A piece of silicon wafer (transfer wafer) is glued to the noble metal film using a thin layer of thermal epoxy (Epo-Tek 375, Epoxy Technology) and then peeled upwards after the epoxy layer achieves its final hardness, resulting in stripping of the gold film. WS$_2$ was mechanically exfoliated from bulk crystal (HQ-graphene) using Scotch Tape and transferred onto the Au substrate. Flakes with different thicknesses (6-30 nm) were transferred onto the substrate.

Patterning of the WS$_2$ flakes was achieved by a combination of electron-beam lithography (Elionix ELS-7500EX) exposure of poly methyl metha acrylate (PMMA A4) and dry etching process. Dry-etching was performed using a combination of XeF$_2$ and Ar for 210 seconds, which is suitable for all the samples with varying thickness from 6 nm to 20 nm to be fully etched. Since XeF$_2$ is highly selective, the etching process does not affect the Au substrate. Finally, the samples were cleaned in acetone to remove the 250 nm-thick PMMA resist. The grating region exhibits a clear change in color compared to the unpatterned region when viewed under an optical microscope in reflection mode under white-light illumination.

**Optical characterization**

The WS$_2$ reflectance spectra were obtained using a microscope with external white-light illumination (AvaLight-HAL) under normal incidence. The reflected light signals were collected from the microscope objective (Olympus SLMPLN 50X N.A.=0.35) and analyzed using a grating spectrometer coupled to a Si focal plane array (FPA). All these instruments are integrated in the LabRAM HR Evolution Confocal Microscope. In order to avoid the signals from the system background and the gold bulk plasmon resonances, the reflectance spectra are normalized by the reflectance of a silver mirror (Thorlabs PF30-03-P01) which can be considered a spectrally uniform high reflector. While the reflectance from the Ag reference mirror is not unity, the minor (few-percent) absorption by the reference mirror will translate to a similarly minor uncertainty in absolute reflectance calibrations. We assume minimum scattering of the incident light due to normal incidence illumination, small sample roughness and very thin features compared to wavelengths. The separated TM and TE polarization signals are obtained using a linear-polarizer (Edmund Linear Glass Polarizing Filter #43-783) for the incident white light. The incident white-light spot size is ~4 µm in diameter (shown in supplementary S1.c).

**Simulations**

Numerical simulations of the electromagnetic response of the WS$_2$ gratings were achieved with the Lumerical FDTD and COMSOL Multiphysics solvers. 1-dimensional heterostructures were solved using the transfer matrix method. The reflectance spectra and field distributions were calculated by simulating incident plane waves with TM or TE polarizations. The anisotropic permittivity of bulk WS$_2$ reported in ref. 5&12 (*5, 12*)was adopted for our simulations.

**Data Availability**

The data that support the conclusions of this study are available from the corresponding author upon request.

**Code Availability**

The Python scripts used in this study for plotting and modeling are available from the corresponding author upon request.

**Acknowledgements**

D.J. acknowledges primary support for this work by the U.S. Army Research Office under contract number W911NF-19-1-0109. D.J. also acknowledges support from Penn Engineering Start-up funds, National Science Foundation (DMR-1905853) and University of Pennsylvania Materials Research Science and Engineering Center (MRSEC) (DMR-1720530) and Northrop Grumman Corporation. A.R.D. acknowledges start-up funds from UCLA. B.A. and D.J. acknowledge support from the Vagelos Institute for Energy Science and Technology at Penn. This work was carried out in its entirety at the Singh Center for Nanotechnology at the University of Pennsylvania which is supported by the National Science Foundation (NSF) National Nanotechnology Coordinated Infrastructure Program grant NNCI-1542153. The authors acknowledge assistance from Xingdu Qiao for SEM imaging.

**Author Contributions**

D.J. and H.Z. conceived the project. H.Z. made the samples, performed microfabrication, reflectance measurements and electron microscopy. B.A. performed all simulations and modelling with help from A.R.D., M.W.K., Q.Z. and S.R. J.M. assisted with sample preparation and microfabrication. K.J. performed atomic force microscopy characterization. D.J., H.Z., B.A. and A.R.D., analyzed and interpreted the optical spectroscopy and simulation data. D.J. supervised the study. All others contributed to writing of the manuscript.

**Competing interests**

The authors declare no competing interests.

**SUPPLEMENTARY MATERIALS**

Supplementary information is available for this paper at:
Correspondence and requests for materials should be addressed to D.J.

**Fig. S1.** Optical images, SEM and AFM profiles of the grating structure

**Fig. S2.** Two-oscillator model fitting of the exciton polariton detuned by thickness

**Fig. S3.** Evolution of the exciton and cavity modes in various material systems

**Fig. S4.** Mode evolution in TE polarization

**Fig. S5. (Movie S5.)** Evolution of SPP modes based on width dependence

**Fig. S6.** Triple-oscillator model calculation and fitting

**Fig. S7.** Evolution of strong coupling as a function of thickness

# REFERENCES AND NOTES

1. J. A. Schuller, E. S. Barnard, W. Cai, Y. C. Jun, J. S. White, M. L. Brongersma, Plasmonics for extreme light concentration and manipulation. *Nature materials* **9**, 193 (2010).
2. B. Luk'yanchuk, N. I. Zheludev, S. A. Maier, N. J. Halas, P. Nordlander, H. Giessen, C. T. Chong, The Fano resonance in plasmonic nanostructures and metamaterials. *Nature materials* **9**, 707 (2010).
3. A. I. Kuznetsov, A. E. Miroshnichenko, M. L. Brongersma, Y. S. Kivshar, B. Luk'yanchuk, Optically resonant dielectric nanostructures. *Science* **354**, aag2472 (2016).
4. L. Zhang, S. Mei, K. Huang, C. W. Qiu, Advances in full control of electromagnetic waves with metasurfaces. *Advanced Optical Materials* **4**, 818-833 (2016).

5. Y. Li, A. Chernikov, X. Zhang, A. Rigosi, H. M. Hill, A. M. van der Zande, D. A. Chenet, E.-M. Shih, J. Hone, T. F. Heinz, Measurement of the optical dielectric function of monolayer transition-metal dichalcogenides: MoS 2, Mo S e 2, WS 2, and WS e 2. *Physical Review B* **90**, 205422 (2014).
6. A. Krasnok, S. Lepeshov, A. Alú, Nanophotonics with 2D transition metal dichalcogenides. *Optics express* **26**, 15972-15994 (2018).
7. D. Zheng, S. Zhang, Q. Deng, M. Kang, P. Nordlander, H. Xu, Manipulating coherent plasmon–exciton interaction in a single silver nanorod on monolayer WSe2. *Nano letters* **17**, 3809-3814 (2017).
8. T. Hu, Y. Wang, L. Wu, L. Zhang, Y. Shan, J. Lu, J. Wang, S. Luo, Z. Zhang, L. Liao, Strong coupling between Tamm plasmon polariton and two dimensional semiconductor excitons. *Applied Physics Letters* **110**, 051101 (2017).
9. J. Wen, H. Wang, W. Wang, Z. Deng, C. Zhuang, Y. Zhang, F. Liu, J. She, J. Chen, H. Chen, Room-temperature strong light–matter interaction with active control in single plasmonic nanorod coupled with two-dimensional atomic crystals. *Nano letters* **17**, 4689-4697 (2017).
10. M. Wang, A. Krasnok, T. Zhang, L. Scarabelli, H. Liu, Z. Wu, L. M. Liz-Marzán, M. Terrones, A. Alù, Y. Zheng, Tunable fano resonance and plasmon–exciton coupling in single au nanotriangles on monolayer WS2 at room temperature. *Advanced Materials* **30**, 1705779 (2018).
11. S. Jahani, Z. Jacob, All-dielectric metamaterials. *Nature nanotechnology* **11**, 23 (2016).
12. R. Verre, D. G. Baranov, B. Munkhbat, J. Cuadra, M. Käll, T. Shegai, Transition metal dichalcogenide nanodisks as high-index dielectric Mie nanoresonators. *Nature nanotechnology*, 1 (2019).
13. D. Jariwala, A. R. Davoyan, G. Tagliabue, M. C. Sherrott, J. Wong, H. A. Atwater, Near-unity absorption in van der Waals semiconductors for ultrathin optoelectronics. *Nano letters* **16**, 5482-5487 (2016).
14. M. A. Kats, R. Blanchard, P. Genevet, F. Capasso, Nanometre optical coatings based on strong interference effects in highly absorbing media. *Nature materials* **12**, 20 (2013).
15. L. C. Flatten, Z. He, D. M. Coles, A. A. Trichet, A. W. Powell, R. A. Taylor, J. H. Warner, J. M. Smith, Room-temperature exciton-polaritons with two-dimensional WS 2. *Scientific reports* **6**, 33134 (2016).
16. S. Pi, X. Zeng, N. Zhang, D. Ji, B. Chen, H. Song, A. Cheney, Y. Xu, S. Jiang, D. Sun, Dielectric-grating-coupled surface plasmon resonance from the back side of the metal film for ultrasensitive sensing. *IEEE Photonics Journal* **8**, 1-7 (2015).
17. U. Schröter, D. Heitmann, Grating couplers for surface plasmons excited on thin metal films in the Kretschmann-Raether configuration. *Physical Review B* **60**, 4992 (1999).
18. X. Zhang, C. De-Eknamkul, J. Gu, A. L. Boehmke, V. M. Menon, J. Khurgin, E. Cubukcu, Guiding of visible photons at the ångström thickness limit. *Nature nanotechnology*, 1-7 (2019).
19. S. Mukherjee, H. Sobhani, J. B. Lassiter, R. Bardhan, P. Nordlander, N. J. Halas, Fanoshells: nanoparticles with built-in Fano resonances. *Nano letters* **10**, 2694-2701 (2010).
20. B. Li, S. Zu, Z. Zhang, L. Zheng, Q. Jiang, B. Du, Y. Luo, Y. Gong, Y. Zhang, F. Lin, Large Rabi splitting obtained in Ag-WS 2 strong-coupling heterostructure with optical microcavity at room temperature. *Opto-Electronic Advances* **2**, 190008 (2019).
21. K.-J. Boller, A. Imamoğlu, S. E. Harris, Observation of electromagnetically induced transparency. *Physical Review Letters* **66**, 2593 (1991).
22. M. Fleischhauer, M. D. Lukin, Dark-state polaritons in electromagnetically induced transparency. *Physical review letters* **84**, 5094 (2000).
23. R. A. Pala, J. S. Liu, E. S. Barnard, D. Askarov, E. C. Garnett, S. Fan, M. L. Brongersma, Optimization of non-periodic plasmonic light-trapping layers for thin-film solar cells. *Nature communications* **4**, 2095 (2013).
24. Z. Li, E. Palacios, S. Butun, K. Aydin, Visible-frequency metasurfaces for broadband anomalous reflection and high-efficiency spectrum splitting. *Nano letters* **15**, 1615-1621 (2015).
25. K. Aydin, V. E. Ferry, R. M. Briggs, H. A. Atwater, Broadband polarization-independent resonant light

26. L. Cao, P. Fan, E. S. Barnard, A. M. Brown, M. L. Brongersma, Tuning the color of silicon nanostructures. *Nano letters* **10**, 2649-2654 (2010).
27. Y. Gu, L. Zhang, J. K. Yang, S. P. Yeo, C.-W. Qiu, Color generation via subwavelength plasmonic nanostructures. *Nanoscale* **7**, 6409-6419 (2015).
28. S. J. Kim, P. Fan, J.-H. Kang, M. L. Brongersma, Creating semiconductor metafilms with designer absorption spectra. *Nature communications* **6**, 7591 (2015).
29. D. Rodrigo, O. Limaj, D. Janner, D. Etezadi, F. J. G. De Abajo, V. Pruneri, H. Altug, Mid-infrared plasmonic biosensing with graphene. *Science* **349**, 165-168 (2015).
30. S. Ren, Q. Tan, J. Zhang, Review on the quantum emitters in two-dimensional materials. *Journal of Semiconductors* **40**, 071903 (2019).
31. Y.-M. He, G. Clark, J. R. Schaibley, Y. He, M.-C. Chen, Y.-J. Wei, X. Ding, Q. Zhang, W. Yao, X. Xu, Single quantum emitters in monolayer semiconductors. *Nature nanotechnology* **10**, 497 (2015).
32. S. Wu, S. Buckley, J. R. Schaibley, L. Feng, J. Yan, D. G. Mandrus, F. Hatami, W. Yao, J. Vučković, A. Majumdar, Monolayer semiconductor nanocavity lasers with ultralow thresholds. *Nature* **520**, 69 (2015).
33. N. Papasimakis, V. A. Fedotov, N. Zheludev, S. Prosvirnin, Metamaterial analog of electromagnetically induced transparency. *Physical Review Letters* **101**, 253903 (2008).
34. H. Deng, H. Haug, Y. Yamamoto, Exciton-polariton bose-einstein condensation. *Reviews of Modern Physics* **82**, 1489 (2010).
35. R. Balili, V. Hartwell, D. Snoke, L. Pfeiffer, K. West, Bose-Einstein condensation of microcavity polaritons in a trap. *Science* **316**, 1007-1010 (2007).
36. H. Deng, G. Weihs, D. Snoke, J. Bloch, Y. Yamamoto, Polariton lasing vs. photon lasing in a semiconductor microcavity. *Proceedings of the National Academy of Sciences* **100**, 15318-15323 (2003).

absorption using ultrathin plasmonic super absorbers. *Nature communications* **2**, 517 (2011).

# Supplementary Information

# Hybrid Exciton-Plasmon-Polaritons in van der Waals Semiconductor Gratings

Huiqin Zhang,[1] Bhaskar Abhiraman,[1,2] Qing Zhang,[3] Jinshui Miao,[1] Kiyoung Jo,[1] Stefano Roccasecca,[2] Mark W. Knight,[4] Artur R. Davoyan,[5] Deep Jariwala[1]*

[1]Department of Electrical and Systems Engineering, University of Pennsylvania, Philadelphia, PA 19104, USA

[2]Department of Physics, University of Pennsylvania, Philadelphia, PA 19104, USA

[3]Department of Electrical and Computer Engineering, National University of Singapore, Singapore 117583, Singapore

[4]Northrop Grumman Corporation 1 Space Park Drive, Redondo Beach, CA 90278, USA

[5]Department of Mechanical and Aerospace Engineering, University of California, Los Angeles, CA 90095, USA

*Corresponding Author: dmj@seas.upenn.edu

## Supplementary Section 1
### Optical images, SEM and AFM profiles of the grating structure

Extended optical images for $WS_2$ grating structures with various thicknesses are shown in Figure S1 a, b, c. Unpatterned $WS_2$ flakes with various colors illustrate different thicknesses. The grating patterns with different width and period sizes (width varies from 200 nm to 300 nm and period varies from 500 nm to 750 nm) on the same flake show clearly the width and period dependence of the reflectance spectral response. The AFM height profile of the specific region in Figure S1.d is shown in Figure S1 e. The sizes of the width (300 nm), period (600 nm) and thickness (15 nm) can be identified through the AFM image and profile. An SEM image with normal incidence is pictured in Figure S1.f. The AFM image (Figure S1.d) and the SEM image (Figure S1.f) show some roughness on the edges of the grating which may explain the small deviations of the experimentally measured reflectance from simulations.

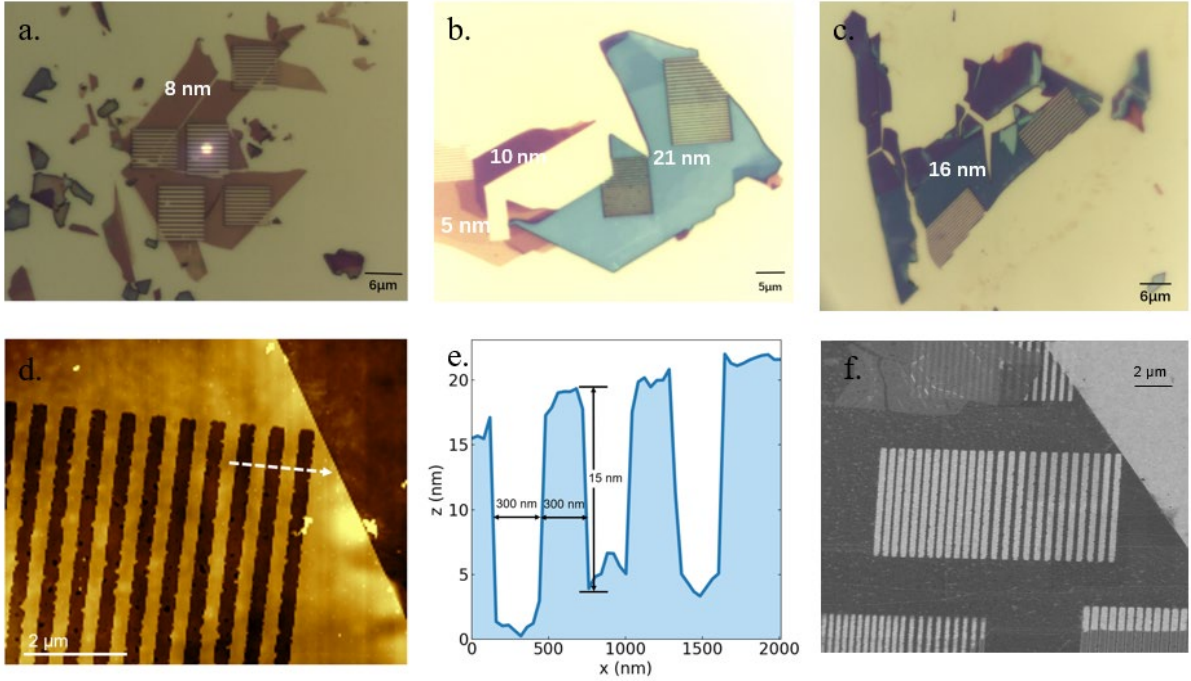

## Supplementary Section 2
**Two-oscillator model fitting of the exciton polariton detuned by thickness(*1-3*)**

The thickness-dependent reflectance spectra clearly show a typical anti-crossing behavior and the formation of the cavity polaritons (Figure 3.a). By fitting the simulation spectra into the calculated coupled two-oscillator model (Figure S2), the evolution of the strong coupling can be obtained by detuning the thickness. The dashed lines in the simulation spectra represent the calculated two-oscillator fitting model which can be represented as:

$\begin{pmatrix} E_C & g_{EC} \\ g_{EC} & E_E \end{pmatrix} \begin{pmatrix} \alpha \\ \beta \end{pmatrix} = E \begin{pmatrix} \alpha \\ \beta \end{pmatrix}$, where $g_{EC}$ represents the coupling strength between the exciton mode and the cavity mode. The exciton mode and cavity mode evolve into the exciton-like polariton and photon-like polariton respectively by strong coupling. At zero detuning, the critical thickness is 15 nm and the Rabi splitting is evaluated as $\hbar\Omega \sim 170$ meV (marked as the arrow in the plot).

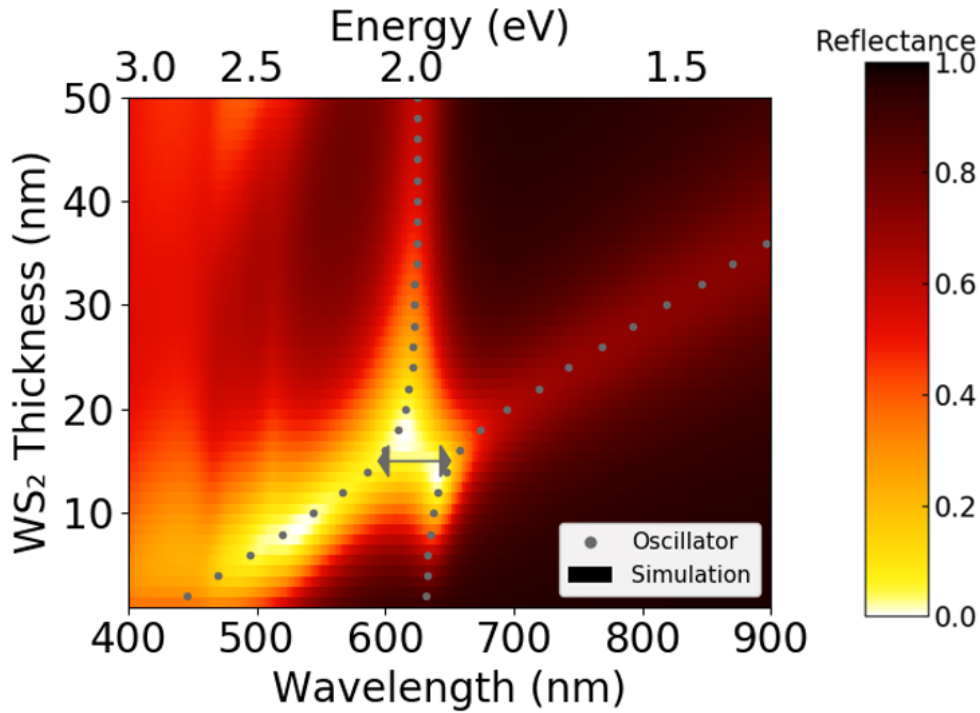

**Supplementary Section 3**

**Evolution of the exciton and cavity modes in various material systems(4)**

Simulated reflection spectra for different dielectric materials ($WS_2$, $WSe_2$, GaP and GaAs) on different plasmonic substrates (Au, Ag and Al) as thickness is swept are shown in Figure S3. The more lossy the plasmonic substrate is (Al>Au>Ag), the higher Fabry-Perot mode resonance absorption the dielectric-metal heterostructure has (in the case of $WS_2$ on Au, Au absorbance averages ~0.2 when $t < t_{cr}$ and ~0.1 when $t > t_{cr}$). In other words, a lossy metal substrate (such as Al) can generate an even stronger cavity mode absorption below the exciton gap. The presence of the high exciton resonance in TMD materials makes $WS_2$ and $WSe_2$ good candidates to demonstrate the strong coupling regime in the visible-light range, compared with normal III-V semiconductors like GaP and GaAs.

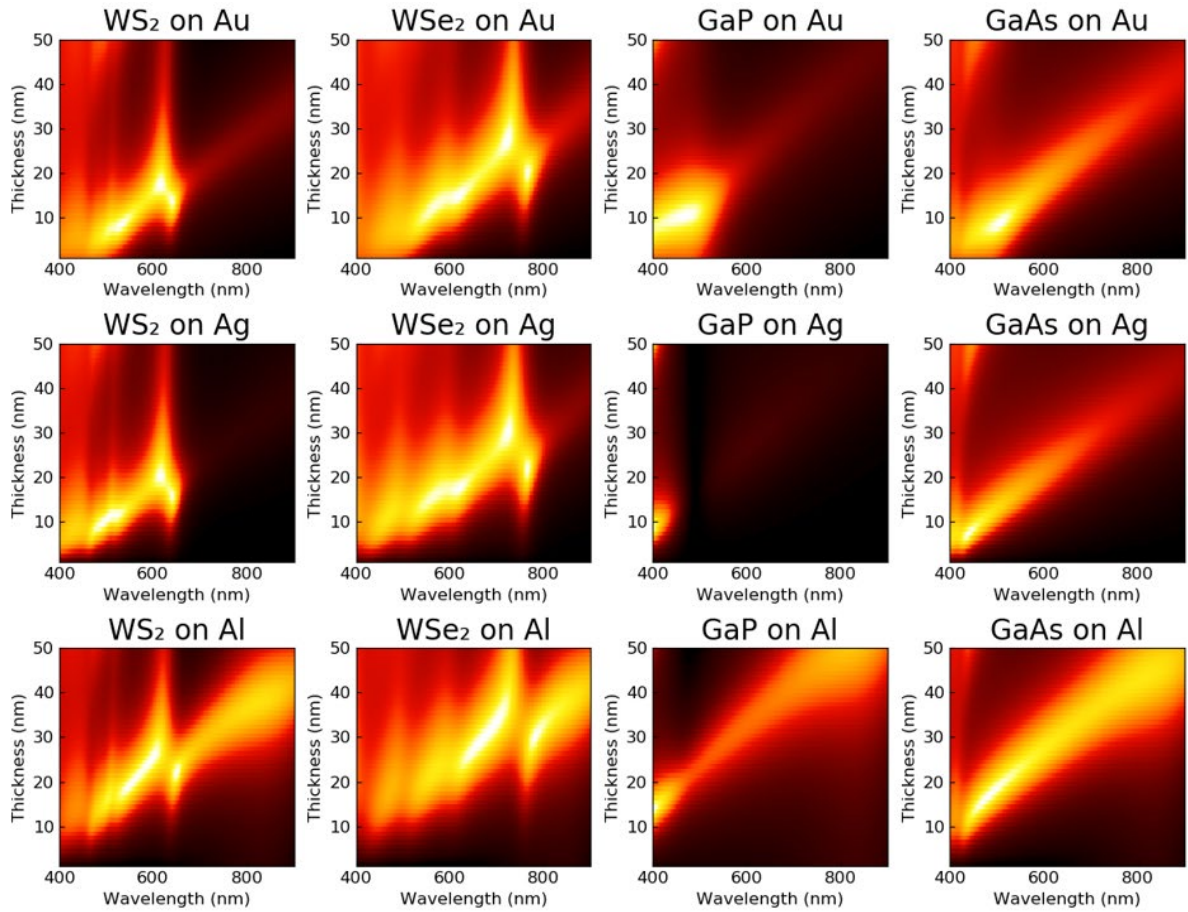

## Supplementary Section 4
### Mode evolution in TE polarization

Simulated reflection spectra in TE polarization as thickness is varied is shown in Figure S4.a. The experimental reflectance spectra with various thicknesses (10 nm, 14 nm, 20 nm) in TE polarization (Figure S4.b) match well with the simulation both qualitatively and quantitatively by the coded circular markers. Figure S4.c shows field profiles of the d-grating mode, SPP mode, and LSPR mode respectively in TE polarization. Here, the $E_y$ field profile in TE polarization corresponds to $H_y$ in TM polarization while $H_z$ in TE polarization corresponds to $E_z$ in TM polarization.

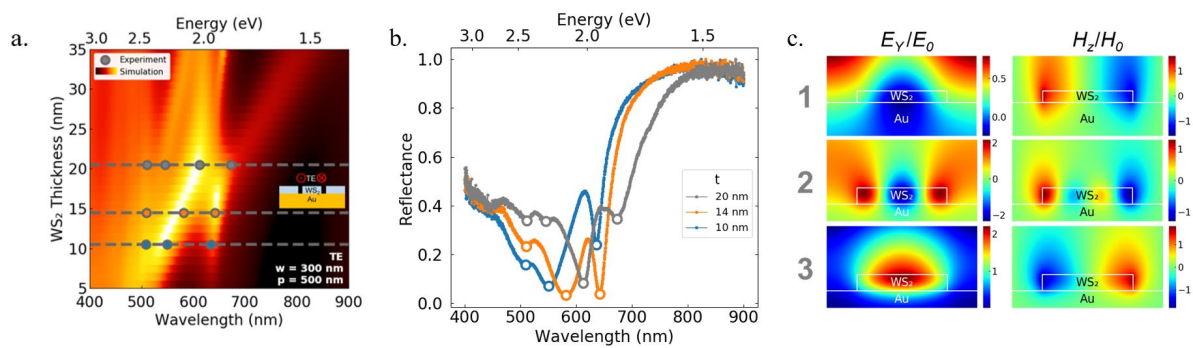

## Supplementary Section 5
### Evolution of SPP modes based on width dependence

The animation from Figure S5.a to i shows the evolution of the SPP peak as the width of $WS_2$ keeps increasing. The SPP branch redshifts and higher-order SPP modes keep emerging as the width is increased (see Movie S5).

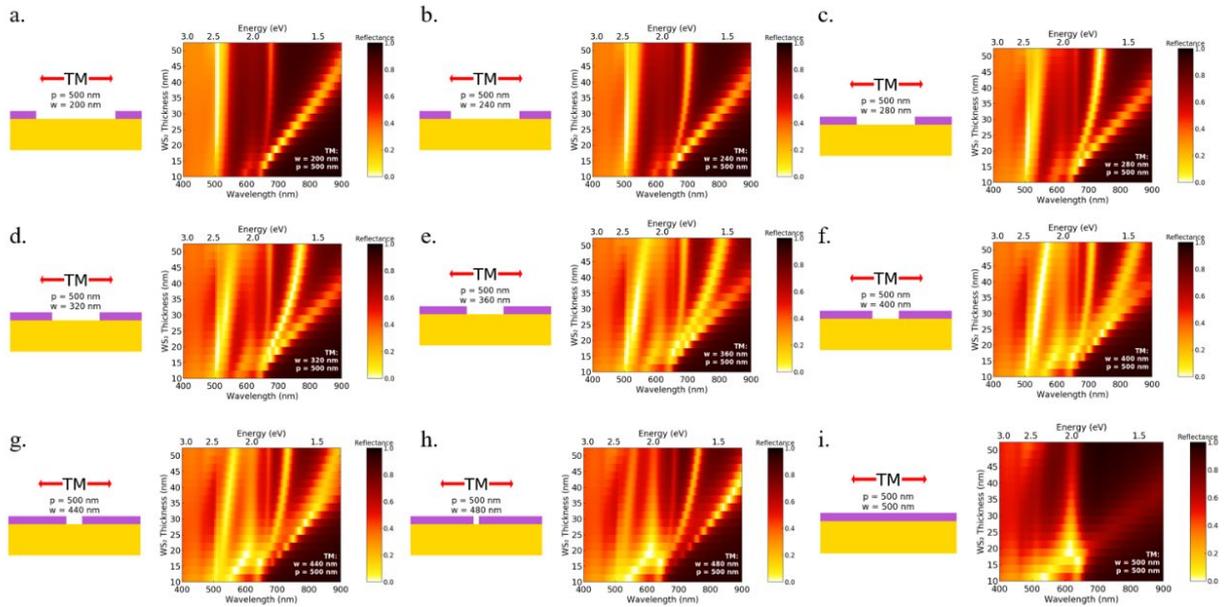

## Supplementary Section 6
### Triple-oscillator model calculation and fitting(5, 6)

The system's absorptive behavior can be modeled classically as a three-particle coupled oscillator system (Figure S6.a). To describe the damped vibration of the linear system with three degrees of freedom which is subjected to an external driven force, the oscillators satisfy the following equation:

$$\boldsymbol{M\ddot{x}} + 2\boldsymbol{G\dot{x}} + \boldsymbol{Kx} = \boldsymbol{F}(t) \text{ with } \boldsymbol{F}(t) = \boldsymbol{F}e^{i\omega t}$$

where M, G, K are real 3x3 symmetric matrices. G and K represent the damping factor and the coupling strength respectively. $\boldsymbol{F}(t) = \begin{pmatrix} F_1 \\ F_2 \\ F_3 \end{pmatrix}$ represents the driven force of each oscillator, just as the mode evolution is driven by light illumination with frequency ω. The period dependence simulation plot (Figure 5.a) is fitted by calculating the frequency response of the hybrid system. To reveal the underlying LEP-UEP-grating coupling behavior in the hybrid nanostructure, the classical coupled oscillator model is employed. Here, oscillator 1, 2 and 3 represent the d-grating mode, the UEP mode and the LEP mode respectively. Here assume,

$$\boldsymbol{M} = \begin{pmatrix} 1 & 0 & 0 \\ 0 & 1 & 0 \\ 0 & 0 & 1 \end{pmatrix}$$

$$\boldsymbol{K} = \begin{pmatrix} k_1 & -k_{12} & k_{13} \\ -k_{12} & k_2 & -k_{23} \\ k_{13} & -k_{23} & k_3 \end{pmatrix} ;$$

Since the d-grating mode's resonant wavelength has a linear dispersion relation with period, $k_1$ has the form of $k_1(p) = a \times (bp + c)^{-2}$ where p is period and a, b, and c are fitting constants. The UEP mode and the EP mode have single-frequency resonances independent of the period.

$$G = W^{-1}\begin{pmatrix} g_1 & 0 & 0 \\ 0 & g_2 & 0 \\ 0 & 0 & g_3 \end{pmatrix} W \qquad W = \frac{1}{2}\begin{pmatrix} 1 & -\sqrt{2} & 1 \\ \sqrt{2} & 0 & -\sqrt{2} \\ 1 & \sqrt{2} & 1 \end{pmatrix}$$

G denotes the damping matrix; $g_1$, $g_2$, and $g_3$ represent the damping factor of each oscillator while W represents a proper rotation matrix. The total energy of the oscillator system is plotted as a function of period and excitation frequency (represented as wavelength) in Figure 5 c. The middle polariton branch (MPB) cut-off we see in electromagnetic simulation and experiment is obtained when only oscillator 2 is driven, i.e. $F_2$ is the only nonzero term in the force vector. Therefore, the MPB mode (corresponding to the UEP mode of the unpatterned case) must be the primary excitation mechanism in this system. By adjusting the fitting parameters in the triple-oscillator model, the calculated response spectra can be obtained as Figure 5.c, a good match with the simulation in Figure 5.a.

The triple coupled harmonic oscillator system can also be represented by transferring the K matrix to the energy matrix:

$$\begin{pmatrix} E_d & g_{d-UEP} & g_{d-LEP} \\ g_{d-UEP} & E_{UEP} & g_{LEP-UEP} \\ g_{d-LEP} & g_{LEP-UEP} & E_{LEP} \end{pmatrix} \begin{pmatrix} \alpha_1 \\ \alpha_2 \\ \alpha_3 \end{pmatrix} = E \begin{pmatrix} \alpha_1 \\ \alpha_2 \\ \alpha_3 \end{pmatrix}$$

To calculate the splitting energy, we assume a near-zero value of $g_{d-LEP}$ so the splitting energy can be expressed in terms of the coupling strengths $g_{d-UEP}$ and $g_{LEP-UEP}$. From the experimental spectrum corresponding to the zero-detuning point (p = 600 nm), $g_{d-UEP}$ and $g_{LEP-UEP}$ are taken to be 116 meV and 89 meV respectively (Figure 5 b). When the d-grating mode and LSPR mode are closest in energy, and the MPB is cut-off in between them, the giant avoided-crossing splitting energy is calculated to be $\hbar\Omega = 2(g_{d-LEP} + g_{LEP-UEP}) = 410 \ meV$. The splitting of 410 meV and the fit from the oscillator model is depicted in k-ω space in Figure S6.b.

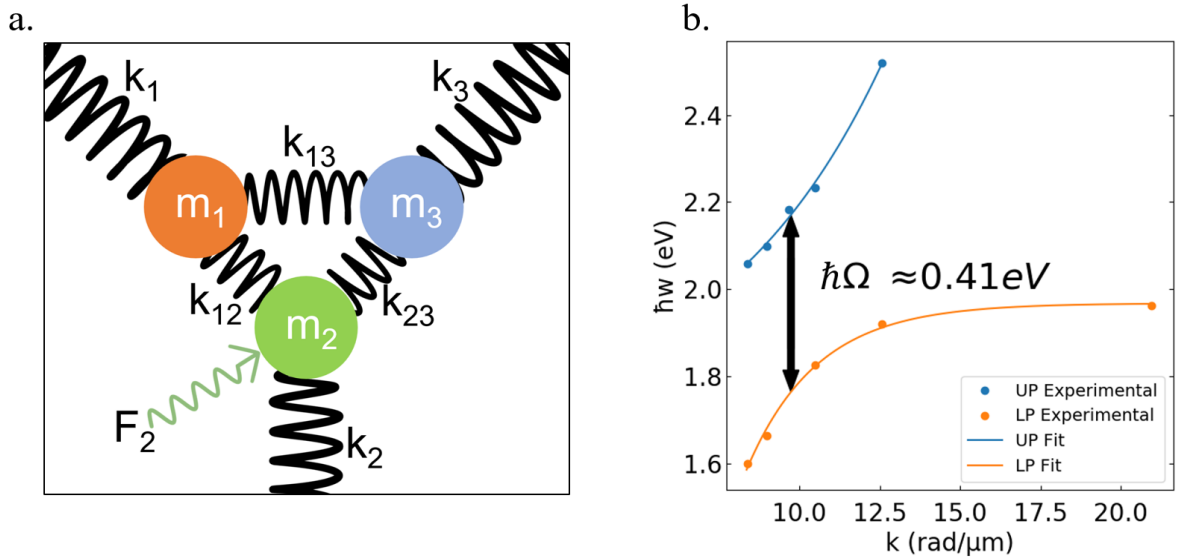

## Supplementary Section 7
### Evolution of strong coupling as a function of thickness
The evolution of the strong coupling as a function of thickness is shown in the experimental reflectance spectra in Figure S7.a. At the zero detuning point where the width equals 300 nm and the period equals 600 nm, the strong coupling evolves with thickness. Reflectance spectra for gratings with fixed w = 300 nm and p = 600 nm and widths of 7 nm and 11 nm are plotted. Superimposing the experimental peaks on the simulation results (Figure S7.b), it is shown that both the d-grating peak (marked as triangles) and the LSPR peak (marked as crosses) redshift as the thickness increases.

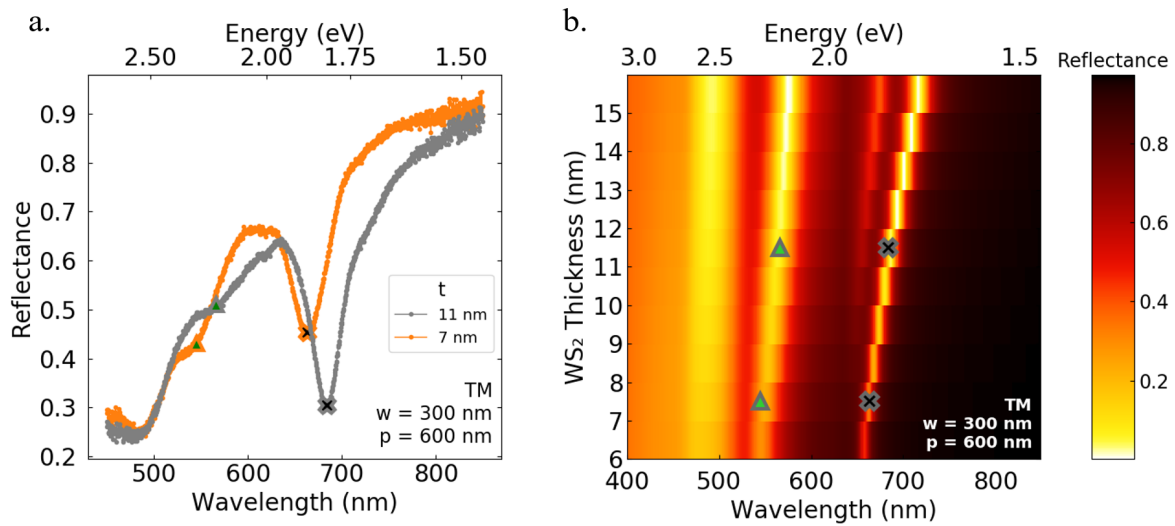

**References:**

1. S. Wang, S. Li, T. Chervy, A. Shalabney, S. Azzini, E. Orgiu, J. A. Hutchison, C. Genet, P. Samorì, T. W. Ebbesen, Coherent coupling of WS2 monolayers with metallic photonic nanostructures at room temperature. *Nano letters* **16**, 4368-4374 (2016).
2. L. Zhang, R. Gogna, W. Burg, E. Tutuc, H. Deng, Photonic-crystal exciton-polaritons in monolayer semiconductors. *Nature communications* **9**, 713 (2018).
3. Q. Wang, L. Sun, B. Zhang, C. Chen, X. Shen, W. Lu, Direct observation of strong light-exciton coupling in thin WS 2 flakes. *Optics express* **24**, 7151-7157 (2016).
4. D. Jariwala, A. R. Davoyan, G. Tagliabue, M. C. Sherrott, J. Wong, H. A. Atwater, Near-unity absorption in van der Waals semiconductors for ultrathin optoelectronics. *Nano letters* **16**, 5482-5487 (2016).
5. C. Qian, S. Wu, F. Song, K. Peng, X. Xie, J. Yang, S. Xiao, M. J. Steer, I. G. Thayne, C. Tang, Two-photon Rabi splitting in a coupled system of a nanocavity and exciton complexes. *Physical review letters* **120**, 213901 (2018).
6. P. Jiang, G. Song, Y. Wang, C. Li, L. Wang, L. Yu, Tunable strong exciton–plasmon–exciton coupling in WS 2–J-aggregates–plasmonic nanocavity. *Optics Express* **27**, 16613-16623 (2019).